\documentclass[12pt,english]{article}
\pdfoutput=1
\usepackage{graphicx,array}
\usepackage{cite}
\usepackage{color}
\usepackage{latexsym}
\usepackage{amsthm}
\usepackage{amsmath}
\usepackage[titletoc]{appendix}
\usepackage{enumitem}
\usepackage{amssymb}
\usepackage[unicode=true,
 bookmarks=true,bookmarksnumbered=false,bookmarksopen=false,
 breaklinks=false,pdfborder={0 0 1},backref=false,colorlinks=true]
 {hyperref}
\hypersetup{pdftitle={What is the Entropy in Entropic Gravity?},
 pdfauthor={Sean M. Carroll and Grant N. Remmen},
 citecolor=blue,linkcolor=blue,urlcolor=blue,citecolor=blue,linkcolor=blue}
\usepackage{breakurl}
\usepackage[hang,flushmargin]{footmisc} 

\def\simgt{\mathrel{\lower2.5pt\vbox{\lineskip=0pt\baselineskip=0pt
           \hbox{$>$}\hbox{$\sim$}}}}
\def\simlt{\mathrel{\lower2.5pt\vbox{\lineskip=0pt\baselineskip=0pt
           \hbox{$<$}\hbox{$\sim$}}}}

\newcommand{\be}{\begin{equation}}
\newcommand{\ee}{\end{equation}}
\newcommand{\bea}{\begin{eqnarray}}
\newcommand{\eea}{\end{eqnarray}}
\newcommand{\Eq}[1]{Eq.~(\ref{#1})}
\newcommand{\Eqs}[2]{Eqs.~(\ref{#1}) and (\ref{#2})}
\newcommand{\Sec}[1]{Sec.~\ref{#1}}

\newcommand{\Fig}[1]{Fig.~\ref{#1}}

\newcommand{\Ref}[1]{Ref.~\cite{#1}}
\newcommand{\Refs}[2]{Refs.~\cite{#1,#2}}
\newcommand{\Tr}{\textrm{Tr\,}}

\newcommand{\HH}{\mathcal{H}}

\newcommand{\ket}[1]{\left| #1 \right\rangle}
\newcommand{\bra}[1]{\left\langle #1 \right |}

\newcommand*\oline[1]{%
  \vbox{%
    \hrule height 0.5pt
    \kern0.68ex
    \hbox{%
      \kern-0.1em
      \ifmmode#1\else\ensuremath{#1}\fi
      \kern-0.1em
    }
  }
}

\definecolor{nicered}{rgb}{0.7,0.1,0.1}
\definecolor{nicegreen}{rgb}{0.1,0.5,0.1}
\hypersetup{
 colorlinks,citecolor=black,linkcolor=black,urlcolor=black}

\begin{document}
\baselineskip=14pt
\hfill CALT-TH-2015-038
\hfill

\vspace{2cm}
\thispagestyle{empty}
\begin{center}
{\LARGE\bf
What is the Entropy in Entropic Gravity?
}\\
\bigskip\vspace{1cm}{
{\large Sean M.\ Carroll and Grant N.\ Remmen}
} \\[7mm]
 {\it Walter Burke Institute for Theoretical Physics,\\
    California Institute of Technology, Pasadena, CA 91125}\let\thefootnote\relax\footnote{e-mail: \url{seancarroll@gmail.com, gremmen@theory.caltech.edu}} \\
 \end{center}
\bigskip
\centerline{\large\bf Abstract}

\begin{quote} \small
We investigate theories in which gravity arises as a consequence of entropy. We distinguish between two approaches to this idea:
 \emph{holographic gravity}, in which Einstein's equation arises from keeping entropy stationary in equilibrium under variations of the geometry and quantum state of a small region, 
and \emph{thermodynamic gravity}, in which Einstein's equation emerges as a local equation of state from constraints on the area of a dynamical lightsheet in a fixed spacetime background. Examining holographic gravity, we argue that its underlying assumptions can be justified in part using recent results on the form of the modular energy in quantum field theory. For thermodynamic gravity, on the other hand, we find that it is difficult to 
formulate a self-consistent definition of the entropy, which represents an obstacle for this approach. This investigation points the way forward in understanding the connections between gravity and entanglement.
\end{quote}
	
\setcounter{footnote}{0}

\newpage
\tableofcontents
\newpage

\section{Introduction}\label{sec:Introduction}

The existence of a profound relationship between gravity and entropy has been recognized since the formulation of the laws of black hole mechanics \cite{BHLaws} and the derivation of the Bekenstein-Hawking entropy \cite{Bekenstein,HawkingBHThermo}. More recently, ideas such as the holographic principle \cite{Holography1,Holography2}, black hole complementarity \cite{Complementarity}, the gauge/gravity correspondence \cite{AdSCFT,Witten,MAGOO}, and the firewall puzzle \cite{AMPS,Braunstein} have provided further hints that a deep relationship between gravitation and entropy will be present in the ultimate theory of quantum gravity.

In the quest to explore this connection and further our understanding of quantum gravity, there have been several proposals for directly linking gravity and entanglement. These proposals fall essentially into two distinct types, which we dub \emph{holographic gravity} (HG) and \emph{thermodynamic gravity} (TG). The labels are not perfect, as HG is related to thermodynamics and TG is related to holography, but they will serve as a useful shorthand for the two approaches.

In holographic gravity, one considers variations of the spacetime geometry and quantum state within a region, posits a relationship between the change in entanglement entropy and the change in the area of the boundary, and
then uses these constraints to derive the Einstein equation in a bulk spacetime. This approach was used successfully in \Refs{Faulkner}{Lashkari} in an AdS/CFT context (see also \Ref{Swingle}) and in \Ref{Jacobson2} in a more general setup based on local causal diamonds. In holographic gravity, gravity emerges as a dual description of the entanglement entropy of the degrees of freedom in a local region.

In thermodynamic gravity, there is no variation over different states. Rather, one fixes a dynamical spacetime and a particular energy-momentum background. One then posits a relationship between some entropy flux (defined using the energy-momentum tensor) and some cross-sectional area (e.g., of a given null surface). Using this area-entropy relation, one derives the Einstein equations. This was the method of \Ref{Jacobson}, as well as Refs.~\cite{Verlinde,Padmanabhan:2009vy,Tian,Cai:2005ra}. While these approaches are similar in spirit, Verlinde \cite{Verlinde} emphasizes the existence of an entropic force from the gradient of the entropy, while Jacobson \cite{Jacobson} derives the Einstein equation directly as a local equation of state.

Open questions are present in both HG and TG approaches. For definiteness, we will focus on Jacobson's version of HG in \Ref{Jacobson2} and of TG in \Ref{Jacobson}. In the HG case, we clarify the underlying assumptions of the theory and present arguments in their favor. In particular, we show how new results on entanglement entropy and the modular Hamiltonian in quantum field theory \cite{Bousso1,Bousso2} can be used to justify a crucial infrared assumption in HG.
On the other hand, we find that TG exhibits a tension related to the fact that the ``entropy'' is not well-defined in this theory. We will argue that it is difficult to find a self-consistent definition of the entropy in TG approaches. Our results indicate that holographic gravity is successful and points the way toward promising future results; reassuringly, holographic gravity is most closely related to AdS/CFT, in that it makes gravity in the bulk of a region dual to entanglement constraints on the boundary, in a sense that we will explore later.

The remainder of this paper is organized as follows. In \Sec{sec:type-A} we first review the holographic formulation of entropic gravity, identify its axioms, and examine its derivation of the Einstein equation. Afterwards, we demonstrate that the axioms of this theory can be justified in part using recent results in quantum field theory. In \Sec{sec:type-B} we examine the formulation of the thermodynamic approach to entropic gravity and demonstrate the origin of the difficulties it experiences in defining the entropy. Finally, we summarize and discuss future directions in \Sec{sec:conclusions}.

\section{Holographic Gravity}\label{sec:type-A}
After reviewing the motivation for relating entropy, particularly that of entanglement, with gravitation, we codify the axioms of holographic gravity and demonstrate the derivation of the Einstein equation. We then investigate how to justify and make rigorous each of the postulates underlying HG.

\subsection{Motivation}\label{sec:Motivation}
We start with the underlying motivation of the holographic approach to entropic gravity. One of the most important facts that we (believe we) know about quantum gravity is the proportionality relationship between entropy $S$ and horizon area $A$ \cite{Hawking75,HawkingBHThermo},\footnote{Throughout, we leave $\hbar$ explicit in expressions leading to the derivation of the Einstein equation, as a bookkeeping device for semiclassicality.}
\be
  S = \frac{A}{4G\hbar}.
  \label{eq:areaformula}
\ee
The derivation of this fact is phenomenological: the energy $E$ of the black hole is given by its mass, Hawking used quantum field theory in curved spacetime to calculate the temperature $T = \hbar/8\pi GM$, and then we can use the thermodynamic relation $1/T = \partial S/\partial E$ to define the entropy. One expects that this entropy represents the logarithm of the underlying degrees of freedom in the true theory of quantum gravity; this expectation has been successfully borne out in certain stringy models of black holes \cite{StromingerVafa,MSW}.

In the black hole case, it is clear what the entropy is actually the entropy of: the black hole, or at least the degrees of freedom that macroscopically appear to us as a black hole. That is a system that can be objectively defined in a way upon which all observers would agree. But the same formula \eqref{eq:areaformula} applies to the horizon of de~Sitter space, as shown by Gibbons and Hawking \cite{GibbonsHawking}. The de~Sitter horizon is an observer-dependent notion; given any worldline extended to future infinity, the horizon separates events within the causal diamond of that worldline from those outside. This suggests that the identification of the entropy as belonging to the system described by the horizon applies more universally than to fixed objects like black holes and indeed may apply to horizons in general.

Another clue comes from the existence of Rindler horizons in Minkowski space. Starting from the vacuum state of a general interacting quantum field theory, the Bisognano-Wichmann theorem guarantees that the density matrix restricted to the wedge $z > |t|$ is that of a thermal state with respect to the boost Hamiltonian \cite{Unruh,BisognanoWichmann}. The boundary of the wedge acts as a horizon for observers who are moving with a constant acceleration along the $z$-axis.  In 3+1 dimensions, the area of this horizon is infinite, so it is unsurprising that the von~Neumann entropy of the corresponding density matrix is also infinite. We can ask, however, about the entropy density per unit horizon area. This is also infinite, which can be attributed to the contributions of ultraviolet modes of the field. Imposing an arbitrary short-distance cutoff, we find that there is a constant, fixed amount of entropy per unit horizon area. Since the original calculation was carried out in flat-spacetime quantum field theory, it is natural to suppose that the true entropy density would be finite in a quantum theory of gravity.\footnote{As noted in \Ref{Jacobson2}, we can say that gravity cuts off the number of degrees of freedom and renders the entropy finite or that demanding a finite horizon entropy implies the existence of gravity. Requiring finite entropy at least implies some ultraviolet cutoff for the applicability of quantum field theory.}

Together, these facts suggest that there is a universal relationship: to any horizon, we can associate an entropy proportional to its area. This observation was the inspiration for entropic gravity in \Refs{Jacobson}{Jacobson2}. It remains to formulate a precise prescription for what kind of entropy is actually involved. The natural candidate in the quantum context
is the von~Neumann entanglement entropy, $-\Tr \rho \log \rho$ for some density matrix $\rho$. Taking a vacuum spacetime region and cutting off modes at some fixed short distance, we obtain an entanglement entropy that is proportional to the area of the boundary of the region being considered \cite{Bombelli,Srednicki}. The entanglement entropy further appears in the recent proofs of versions of the covariant entropy bound within quantum field theory \cite{Bousso1,Bousso2}. Moreover, the Ryu-Takayanagi formula \cite{RT,LewkowyczMaldacena} in AdS/CFT relates the entanglement entropy in a boundary region with the area of an extremal surface in the bulk. The conjectured ER=EPR duality \cite{EREPR} (see also \Refs{BPR1}{BPR2}) further underscores this connection. Taking these results as motivation, holographic gravity seeks to relate the Einstein equations themselves to constraints on entanglement entropy and areas in a sense that we will make precise.

\subsection{Formulation of holographic gravity}\label{sec:HGformulation}

\begin{figure}[t]
\begin{center}
\includegraphics[width=.55\textwidth]{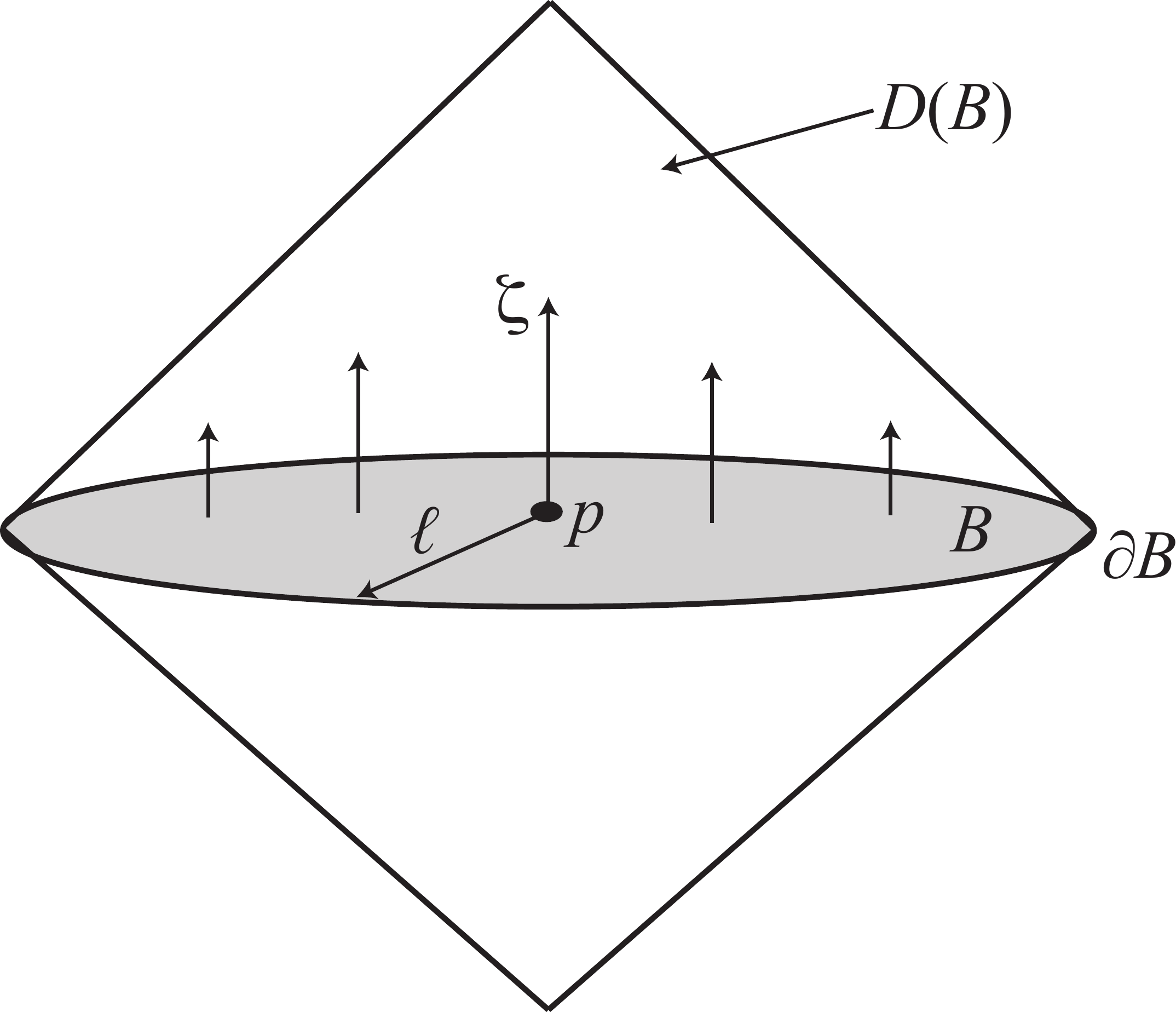}
\end{center}
\caption{A small causal diamond $D(B)$ for a spacelike ball $B$ with boundary $\partial B$. The ball is defined as all points in some spacelike surface that are less than or equal to a distance $\ell$ from some point $p$. The vector field $\zeta$ generates a conformal isometry within $D(B)$, assumed to be approximated by a maximally symmetric spacetime.}
\label{fig:causaldiamond}
\end{figure}

Let us now review the approach to HG laid out in \Ref{Jacobson2}. Fix an arbitrary background $D$-dimensional spacetime geometry $M$ and a spacelike slice $\Sigma$. Choose a point $p\in\Sigma$ and define a ball $B$ as the set of points $p'\in\Sigma$ such that the geodesic distance in $\Sigma$ between $p$ and $p'$ is less than $\ell$. Next, define the {\it causal diamond} $D(B)$ associated with $B$ as the union of the past and future domains of dependence of $B$; that is, the set of all points $x \in M$ such that all inextendible timelike curves through $x$ necessarily intersect $B$; see \Fig{fig:causaldiamond}. We write as $V$ the volume of $B$ and $A$ the area of $\partial B$. For a sufficiently small causal diamond, the background metric approaches the Minkowski form. There is a unique conformal isometry generated by the Killing vector 
\be
\zeta = \frac{1}{2\ell}[(\ell^2 - u^2)\partial_u + (\ell^2 - v^2)\partial_v],\label{eq:zeta}
\ee
where $u=t-r$ and $v=t+r$ for time coordinate $t$ and radial coordinate $r$. 

Writing the quantum state of the system on $\Sigma$ as $\ket{\psi}$, we can define the reduced density matrix on $B$ as 
\be 
\rho_B = \Tr_{\Sigma-B} \ket{\psi}\bra{\psi}.
\ee
We define the entanglement entropy associated with $B$ as
\be
S_{B} = -\Tr \rho_B \,\log\rho_B,
\ee
i.e., the entanglement of the state on $B$ with that on $\Sigma-B$. We posit that the Hilbert space of states on $B$ can be factorized into infrared and ultraviolet contributions, 
\be
\HH_B = \HH_{\rm UV}\otimes\HH_{\rm IR}. 
\ee
The infrared states are ordinary field-theory states in a spacetime background (including semiclassical gravitational perturbations), while the ultraviolet contributions represent short-distance physics, including specifically quantum-gravitational degrees of freedom. Writing $\Lambda_{\rm UV}$ for the scale of the UV completion, which we take to be below the Planck scale, then $\HH_{\rm IR}$ and $\HH_{\rm UV}$ contain degrees of freedom with energies below and above $\Lambda_{\rm UV}$, respectively. The size $\ell$ of the causal diamond is taken to be larger than the Planck length but smaller than $1/\Lambda_{\rm UV}$.
Tracing out the UV degrees of freedom, we are left with an infrared density matrix 
\be
\rho_{\rm IR} = \Tr_{\rm UV}\rho_B.  
\ee
We  then define the (field-theoretic) modular Hamiltonian $K$ on $B$ via the implicit relation
\be
\rho_{\rm IR} = \frac{e^{-K}}{\Tr e^{-K}}. \label{eq:modHam}
\ee

In Minkowski space, the causal diamond $D(B)$ can be mapped via a conformal transformation to the Rindler wedge \cite{CasiniHuertaMyers}; writing $x^\mu = (t,r,\vec{y})$ for the (radial) coordinates of the ball, $X^\mu = (X^0,X^1,\vec{Y})$ for the (Cartesian) coordinates of the Rindler wedge $X^1>0$, and defining $B^\mu =  (0,1,0,...,0)/2\ell$, the conformal transformation is
\be
x^\mu = \frac{X^\mu - B^\mu X^2}{1 - 2X\cdot B + B^2 X^2} + 2 \ell^2 B^\mu,\label{eq:conformaltransformation}
\ee 
where $X^2 = X^\mu X_\mu$ and similarly for $B^2$. With $U=X^0-X^1$ and $V=X^0+X^1$, the Rindler wedge corresponds to the intersection of $V>0$ and $U<0$, which maps to the causal diamond $D(B)=\{v<\ell\}\cap\{u>-\ell\}$. 
For the Rindler wedge, the Bisognano-Wichmann theorem \cite{BisognanoWichmann} guarantees that the density matrix is thermal with respect to the Hamiltonian generating time translation. Thus, for a conformal field theory (CFT), which is invariant under this transformation of the geometry, the modular Hamiltonian $K$ is just the Hamiltonian generating flow along $\zeta$ from \Eq{eq:zeta}, namely,
\be 
K_{\rm CFT} = \frac{2\pi}{\hbar}\int_B T^{\mu\nu} \zeta_\mu \mathrm{d}\Sigma_\nu,\label{eq:KCFT}
\ee
where $\mathrm{d}\Sigma_\mu$ is the surface element orthogonal to $B$ and $T_{\mu\nu}$ is the energy-momentum tensor.

We now consider a variation of the spacetime $M$ and of the quantum state $\rho_B$ on $B$. We will write this variation via
\be 
\delta_{g,\rho}: \text{variation of state $\rho_B$ and geometry $g$ that keeps the volume $V$ of $B$ fixed.}
\ee 
Under this variation, the area $A$ at fixed $V$ changes by $\delta_{g,\rho} A|_V$ and the quantum state $\rho_B$ on $B$ changes by $\delta_{g,\rho} \rho_B$. Moreover, there is a change in the entanglement entropy $\delta_{g,\rho} S_{B}$ as well as a change in the expectation value of the modular Hamiltonian, $\delta_{g,\rho} \langle K \rangle$, which is in general a highly nonlocal quantity. The modular Hamiltonian does not correspond {\it a priori} to any intuitive sense of energy; it is just an operator one can define using the reduced density matrix. Note that all of the above variations are {\it not} dynamical variations that occur with time; rather, we are considering varying the entire history of the configuration, examining the consequences for various quantities for infinitesimally separated configurations of geometry and fields. For example, for a CFT, plugging in the Killing field \eqref{eq:zeta} into \Eq{eq:KCFT} and requiring a sufficiently small causal diamond $\ell\ll L_T$, where $L_T$ is the characteristic length scale of changes in $T_{\mu\nu}$, we have the modular energy
\be
\delta_{g,\rho} \langle K_{\rm CFT} \rangle = \frac{2\pi}{\hbar} \frac{\Omega_{D-2}\ell^D}{D^2 - 1}\delta_{g,\rho} \langle T_{00} \rangle,\label{eq:dKCFT}
\ee
where $\Omega_{D-2}=2\pi^{(D-1)/2}/\Gamma[(D-1)/2]$ is the area of the unit $(D-2)$-sphere.

We are now ready to state the postulates of the holographic gravity theory given in \Ref{Jacobson2}. They are as follows:
\begin{enumerate}
\item {\bf Entanglement separability.} The entropy $S_{B}$ can be written as a simple sum $S_{\rm UV} + S_{\rm IR}$, where UV and IR denote the entanglement entropies in the UV (quantum gravitational) and IR (quantum field-theoretic) degrees of freedom. 
Equivalently, the quantum mutual information $I_B = S_{\rm UV} + S_{\rm IR} - S_{B}$ is negligible.
That is, there is minimal entanglement among degrees of freedom at widely separated energy scales.\footnote{This formulation of postulate 1. is actually somewhat stronger than necessary; for holographic gravity it is sufficient that merely the entropy variation $\delta_{g,\rho} S_{B}$ factorize as in \Eq{eq:dSfactor}. However, the justification for this weaker version of postulate 1. will ultimately be the same as the stronger version we state above.}

\item {\bf Equilibrium condition.} The entanglement entropy of the causal diamond is stationary with respect to variations of the state and metric, i.e.,
\be 
\delta_{g,\rho} S_{B} = \delta_{g,\rho} S_{\rm UV} + \delta_{g,\rho} S_{\rm IR} = 0,\label{eq:dSfactor}
\ee 
and the geometry of the causal diamond is that of a maximally symmetric spacetime (Minkowski, de Sitter, or anti-de Sitter).

\item {\bf Area-entropy relation.} The variation of the UV entropy of the causal diamond is proportional to its area change at fixed volume, 
\be 
\delta_{g,\rho} S_{\rm UV} = \eta\, \delta_{g,\rho} A|_V,
\ee 
for some universal constant $\eta$. That is, $\delta S$ satisfies a local, bulk version of holography. This is Jacobson's generalization of the area law for black hole entropy and is the crucial substantive assumption underlying holographic gravity.

\item {\bf Modular energy: CFT form.} The modular energy, defined to be the variation in the expectation value of the modular Hamiltonian, for an arbitrary quantum field theory is given by the form in \Eq{eq:dKCFT}, possibly modified by some scalar operator $X$,
\be
\delta_{g,\rho} \langle K \rangle = \frac{2\pi}{\hbar}\frac{\Omega_{D-2}\ell^D}{D^2 - 1}\delta_{g,\rho}\left(\langle T_{00} \rangle + \langle X \rangle g_{00} \right).\label{eq:post4}
\ee

\end{enumerate}
While the first three postulates are assumptions about ultraviolet behavior, the fourth is strictly an infrared statement and we will argue that, in its null-limit form, it can be derived rather than postulated.
Note that in postulate 3., we expect $\eta=1/4G\hbar$, the same constant as appears in the Bekenstein-Hawking formula \cite{HawkingBHThermo}. 

Reference~\cite{Jacobson2} shows how postulates 1. through 4. can be used to derive the Einstein equations. Our purpose in this section is to illustrate how some of these postulates can be justified rigorously, rather than taken as assumptions. While we leave the geometric details of how the postulates imply the Einstein equations to \Ref{Jacobson2}, we sketch the main points. 
First, writing as usual
\be
S_{\rm IR} =  -\Tr \rho_{\rm IR} \log \rho_{\rm IR},
\ee
we have the entanglement first law \cite{Blanco}
\be 
\delta_{g,\rho} S_{\rm IR} = -\Tr [(\delta_{g,\rho} \rho_{\rm IR}) \log \rho_{\rm IR}] - \Tr(\rho_{\rm IR} \rho_{\rm IR}^{-1} \delta_{g,\rho} \rho_{\rm IR}) = \Tr(K\delta_{g,\rho} \rho_{\rm IR}) = \delta_{g,\rho} \langle K \rangle.
\label{eq:firstlaw}
\ee
Further, the area variation at constant $V$ for a maximally symmetric spacetime in which $G_{\mu\nu} = -f\, g_{\mu\nu}$ for some arbitrary constant $f$ is
\be
\delta_{g,\rho} A|_V = -\frac{\Omega_{D-2} \ell^D}{D^2 - 1}(G_{00} + f\, g_{00}).\label{eq:dAlambda}
\ee
Equating \Eqs{eq:firstlaw}{eq:post4} via postulate 4., setting \Eq{eq:dAlambda} to $\delta S_{\rm UV}/\eta$ via postulate 3., and then putting everything together via postulates 1. and 2., we have
\be 
0 = \delta_{g,\rho} S_{B} = \frac{\Omega_{D-2}\ell^D}{D^2 - 1}\left[-\eta(G_{00} + f\, g_{00}) + \frac{2\pi}{\hbar} \delta_{g,\rho}\left(\langle T_{00} \rangle + \langle X \rangle g_{00} \right)\right].\label{eq:dStot}
\ee
Rearranging and requiring that this relation hold for all possible spatial slicings (i.e., in arbitrary reference frames) requires
\be
R_{\mu\nu} -\frac{1}{2} R g_{\mu\nu} + f\, g_{\mu\nu} = \frac{2\pi}{\hbar\eta}\delta_{g,\rho}\left(\langle T_{\mu\nu} \rangle + \langle X \rangle g_{\mu\nu} \right).
\ee
Now, since we must have $\nabla^\mu T_{\mu\nu} = 0$ for energy-momentum conservation, but $\nabla^\mu R_{\mu\nu} = \nabla_\nu R/2$ by the Bianchi identity, $f$ can be identified as $2\pi \delta_{g,\rho}\langle X \rangle/\hbar\eta + \Lambda$ for arbitrary constant $\Lambda$, yielding Einstein's equation in semiclassical terms,
\be
R_{\mu\nu} - \frac{1}{2} R g_{\mu\nu} + \Lambda g_{\mu\nu} = \frac{2\pi}{\hbar\eta} \, \delta_{g,\rho} \langle T_{\mu\nu} \rangle = 8 \pi G \, \delta_{g,\rho} \langle T_{\mu\nu} \rangle,
\ee
where in the final equality we plugged in $\eta=1/4G\hbar$ as expected for consistency with the Bekenstein-Hawking formula. Note that the $\delta_{g,\rho} \langle T_{\mu\nu} \rangle$ appearing on the right-hand side is really just the expectation value of the energy-momentum tensor under consideration, since without the variation, i.e., in vacuum, the causal diamond is assumed to be described by a maximally symmetric spacetime with vanishing $T_{\mu\nu}$.

The way in which the Einstein equation arose in the above derivation was by the imposition of a relationship between the change of entanglement entropy and area for variations over the spacetime configuration and quantum state. It is not a dynamical constraint within a single solution, but rather a relationship between infinitesimally separated spacetime histories and geometries. Mathematically, how this constraint leads to the Einstein equation is  the same as how the Einstein equation was derived \cite{Faulkner,Lashkari} in the context of AdS/CFT via the Ryu-Takayanagi formula \cite{RT,LewkowyczMaldacena}. That is, AdS/CFT itself, in \Refs{Faulkner}{Lashkari}, provides another realization of holographic gravity. The version of the theory in \Ref{Jacobson2} attempts wider applicability, by applying holographic formulas to causal diamonds in an arbitrary spacetime. It is therefore crucial to investigate the extent to which the postulates of the theory can be justified. We conduct such an investigation in the next subsection, providing a nontrivial check of the health of HG.

\subsection{Justifying the assumptions of holographic gravity}\label{sec:HGjustify}

Postulates 1. through 3. above deal with the ultraviolet degrees of freedom in the ultimate theory of quantum gravity. Hence, they either must be taken as axioms of the theory or shown to be true in a more general ultraviolet completion of gravity (e.g., through holography and string theory). Despite this ultraviolet character, there are motivations for postulates 1. through 3., which we will briefly mention. More importantly, we offer a derivation of a null-limit version of postulate 4., allowing it to be removed as an independent assumption in HG.

Postulate 1., requiring minimal entanglement between infrared and ultraviolet degrees of freedom, is a basic feature of effective field theory \cite{Decoupling}, so the first postulate amounts to the assertion that effective field theory is (at least approximately) valid for the field-theoretic degrees of freedom. That is, for renormalization group flow to work in the usual manner, we require a decoupling between the low- and high-momentum states. We do not expect significant mutual information between the low-energy degrees of freedom in a Wilsonian effective action and those in the ultraviolet completion. This was explicitly found to be the case for interacting scalar quantum field theories in \Ref{Decoupling}.

Postulate 2. is really the entropic foundation of the theory, being the
assertion of a condition on the spacetime geometry that will ultimately
lead to the Einstein equation. In essence, postulate 2. is the assertion
that the vacuum should look as simple as possible, namely, that a small
region should be well described by a Gibbs state. For a fixed energy
expectation value, the Gibbs state has the maximum entropy, so $\delta
S_{B} = 0$. Moreover, for the Gibbs distribution, expectation values
of quantum mechanical quantities related to the entanglement entropy map
onto those from classical thermodynamics \cite{vNQMBook}. Viewed in this sense, the
causal diamond represents a canonical ensemble \cite{Jacobson2}, with fixed degrees of
freedom and volume. Hence, classically, its entropy for a given
expectation value of $T_{\mu\nu}$ is maximized in equilibrium. 
The requirement that the causal
diamond be described by a maximally symmetric spacetime means that there is not power in
spacetime fluctuations at arbitrarily small scales. If this were not the
case, then introducing fluctuations would produce a large backreaction that
would spoil the equilibrium condition. The content of postulate~2. is
therefore the assertion that the semiclassical Einstein equations hold if
and only if the causal diamond is in thermodynamic equilibrium.

Postulate 3. is related to the Ryu-Takayanagi relation \cite{RT,LewkowyczMaldacena}, with which \Refs{Faulkner}{Lashkari} derived the Einstein equations in a holographic context in a manner closely related to that of \Ref{Jacobson2}. References~\cite{Faulkner,Lashkari} can be regarded as another example of HG, in which bulk gravitation is again found to be dual to a constraint on entanglement entropy in some boundary degrees of freedom. The boundaries of the causal diamond can be viewed as the Rindler horizons of a set of appropriately accelerating observers. The area of $\partial B$ is just the area of this horizon. Postulate 3. does not require assigning a change in entropy with time to a dynamical change in area. Rather, it just requires identifying the area of the causal diamond with the entanglement entropy and then doing this for an entire family of infinitesimally separated causal diamond configurations. The motivations for assigning an entropy to an area for this apparent horizon in the first place were discussed in \Sec{sec:Motivation}.

Postulate 4. is of a different character. Unlike the ultraviolet-dependent postulates 1. through 3., postulate 4. is an assertion about the form of the modular Hamiltonian for the field-theoretic degrees of freedom. Thus, postulate 4. is amenable to analysis and, as a consistency test of the holographic gravity of \Ref{Jacobson2}, we can investigate whether postulate 4. can be justified, rather than taken as an assumption. A holographic justification of postulate 4. for spacelike slicing was
considered in \Ref{NewCasini}, in which the subtleties of the
construction in \Ref{Jacobson2} for operators of particular conformal
dimensions is discussed in detail. However, we will show that postulate 4.
may be justified more simply in the null limit by using the conformal symmetry of the causal
diamond and the light-sheet results of \Ref{Bousso2}.

{\it A priori}, postulate 4. suffers from two potential weaknesses. First, it is unclear why the modular energy of a generic quantum field theory should take the form appropriate for a CFT. Second, \Ref{Jacobson2} derived the Einstein equations only for small variations about the vacuum in the field-theoretic density matrix $\rho_{\rm IR}$, for which the entanglement first law \eqref{eq:firstlaw} holds. However, small variations to the geometry, in which gravitational backreaction is negligible, do not necessarily correspond to small variations in $\rho_{\rm IR}$. For example, two massive particles in a Bell pair state certainly gravitate, but their long-range entanglement does not correspond to a small perturbation about the vacuum state of $\rho_{\rm IR}$. Thus, the question remains of how to retain the success of HG in obtaining the Einstein equation for large changes to the quantum state without using the entanglement first law. Both of these challenges can be addressed using recent results proven in quantum field theory. 

To address the second issue, we consider the computation of the modular energy and entanglement entropy for an interacting CFT in $D>2$, which was computed for a null slab in \Refs{Bousso1}{Bousso2}. For an arbitrary state $\rho_{\rm IR}$ defined on a spatial region (for example, one of the spatial slices of our causal diamond), we can define the Casini entropy
\be
\Delta S = -\Tr \rho_{\rm IR} \log \rho_{\rm IR} + \Tr \sigma_{\rm IR} \log \sigma_{\rm IR}, \label{eq:Casini}
\ee
which is just the vacuum-subtracted von~Neumann entropy, and the modular energy,
\be 
\Delta K = \Tr K \rho_{\rm IR} - \Tr K \sigma_{\rm IR},\label{eq:DeltaK}
\ee
where the modular Hamiltonian $K$ is defined as in \Eq{eq:modHam} but with respect to the vacuum density matrix $\sigma_{\rm IR}$. 
Note that in the limit in which the field-theoretic density matrix for this state is infinitesimally close to the vacuum state $\sigma_{\rm IR}$, we have $\Delta S \rightarrow \delta S$ and $\Delta K \rightarrow \delta K$ and the entanglement first law guarantees $\delta S = \delta K$.
Using the replica trick \cite{Replica1,Replica2} to compute the $n$th R{\' e}nyi entropy for an arbitrary spatial region by inserting defect operators on the boundaries, \Ref{Bousso2} shows through an argument involving the operator product expansion in the null limit that the only operators that can contribute to $\Delta K$ or $\Delta S$ are single-copy scalar operators with twist $\tau$ in the range 
\be
\frac{1}{2}(D-2) < \tau \leq D-2.
\ee
For spin-zero operators, $\tau$ is just the scaling dimension. By single-copy, we mean that the operator appears inside just one of the copies of the CFT in the replica trick; in that case, the contribution of this operator to the entanglement entropy is proportional to the expectation value of the operator inside a single copy of the CFT \cite{Bousso2}. That is, single-copy operators contribute linearly in the density matrix to $S_{\rm IR}$. 

Finally, the modular Hamiltonian is the unique operator on $B$ that matches $S_{\rm IR}$ at linear order for arbitrary perturbations of the density matrix. Thus, single-copy operators contribute equally to $\Delta S$ and $\Delta K$, so taking the null limit of any spatial surface and computing $\Delta S$ and $\Delta K$, we have
\be
\Delta S = \Delta K \qquad [\text{null limit}].\label{eq:DSDK}
\ee
One can show that, evaluated on any fixed spatial slice, $\Delta K - \Delta S = D(\rho_{\rm IR}|\sigma_{\rm IR})$, the relative entropy between the state and the vacuum, which is always non-negative. However, in the null limit, \Ref{Bousso2} showed that in an interacting conformal field theory, no operators in the algebra can be localized to a null surface, which allows the excited state and the vacuum to differ while remaining indistinguishable. Moreover, the null limit is sensitive only to the UV structure of the theory. For quantum field theories with an interacting UV fixed point, \Ref{Bousso2} thus showed that the $\Delta S = \Delta K$ result of \Eq{eq:DSDK} continues to hold (provided the quantum field theory does not have finite wave function renormalization, as for, e.g., superrenormalizable theories). The result is therefore quite general.
Equation~\eqref{eq:DSDK} applies to {\it any} quantum state that backreacts weakly on the geometry and thus strengthens the argument for HG in \Ref{Jacobson2}. No longer is it necessary to rely on the entanglement first law \eqref{eq:firstlaw} and consider only small perturbations about the vacuum density matrix in postulate 4.; we are now free to consider arbitrary states.

To mitigate the first issue raised regarding postulate 4., namely the question of why the modular energy for a generic quantum field theory should be related to that of a CFT, \Ref{Jacobson2} considered a quantum field theory with a UV fixed point and required that the size of the causal diamond be smaller than every length scale in the quantum field theory, i.e.,
\be
\ell \ll \frac{1}{\max_i m_i},\label{eq:smalldiamond}
\ee
where $m_i$ are the masses of states in the quantum field theory. That is, we are required to take the causal diamond to be smaller than the cutoff of the quantum field theory, $\Lambda_{\rm UV}\gg\ell$. Naively, this leads to doubtful consistency of treating the spacetime semiclassically; we do not want to be required to take the causal diamond to be Planck-scale. However, we typically expect the scale of a perturbative UV completion of gravity to be parametrically smaller than the Planck scale, as indeed is the case in string theory \cite{WGC}. 

In any case, we can dramatically relax the stipulation of \Eq{eq:smalldiamond} by evaluating $\Delta K$ in the null limit. Let us choose a sequence of spacelike slices $B_\xi$ through the diamond, $\xi \in [0,1]$, defined by the orbit of $\zeta$ in \Eq{eq:zeta}, where we start with $B_0 = B$ and end with  $B_1$, the upper null surface of the diamond. Now, $\zeta$ is not a member of the Poincar\'e group; it is a conformal Killing vector and in particular contains a dilation. The proper distance across $B_\xi$ tends to zero as we send $\xi\rightarrow 1$, that is, as we flow along $\zeta$. Acting with $\zeta$ on a given field configuration with small momentum on $B_0$ takes us to larger and larger momenta and we experience renormalization group flow as we move through different values of $\xi$. For a CFT, $\zeta$ acts trivially, but for a general interacting quantum field theory with a UV fixed point \cite{Jacobson2,Bousso2}, flow along $\zeta$ means are probing higher and higher energy scales within the theory, eventually reaching a regime in which the CFT approximation is valid. Thus, \Eqs{eq:KCFT}{eq:post4} become in the null limit
\be 
\Delta K = \frac{2\pi}{\hbar} \frac{\Omega_{D-2} \ell^D}{D^2 - 1}T_{uu},
\ee
where we still assume that $T_{\mu\nu}$ varies with a length scale $L_T$ larger than $\ell$. Moreover, we can write the area variation \eqref{eq:dAlambda} in terms of its null components as $\zeta$ lines up with $\partial_u$ in the null limit. Hence, \Eq{eq:dStot} still applies, but for the $uu$ components. Following the logic through as before, we again obtain the Einstein equation
\be
G_{\mu\nu} + \Lambda g_{\mu\nu} = 8\pi G T_{\mu\nu}. 
\ee

We therefore see the infrared assumption underlying holographic gravity, that the modular energy takes the CFT form given in \Eq{eq:post4}, need not be separately postulated, but can be justified by examining the null limit. The null surfaces themselves are not special or preferred, but the use of the null limit rendered tractable the explicit computation of the entanglement entropy and modular energy for generic interacting quantum field theories with an ultraviolet fixed point. Moreover, we seem to have a specific and self-consistent formulation of what kind of entropy we are talking about in holographic gravity: the Casini entropy evaluated on the null boundary of a small causal diamond.

\section{Thermodynamic Gravity}\label{sec:type-B}

In \Sec{sec:type-A}, we demonstrated that the holographic gravity of \Ref{Jacobson2} can be made well defined, putting its axioms on a more solid footing. In this section, we turn to the question of whether the same can be done for the thermodynamic gravity of \Ref{Jacobson}. We will argue that there does not exist any self-consistent definition of entropy in this approach.

In the cases relating entropy and area that we discussed in \Sec{sec:Motivation}, the area is a constant along the horizon. To work our way toward a truly dynamical theory of gravity, we must be able to handle more general cases, including time-dependent spacetimes. Holographic gravity accommodates this requirement by varying the spacetime history: in that case, general relativity can be shown to be equivalent to constraints relating the variation in entanglement entropy and area of a small causal diamond. However, the HG approach does not allow gravity to truly emerge as an equation of state, since the area and entropy variations are not dynamical changes within a single background spacetime. TG takes the other approach. That is, we can start with the null generators of a local Rindler horizon, but the corresponding cross-sectional area will generally change with time. Thermodynamic gravity \cite{Jacobson} therefore posits that the \emph{change} in entropy behind such a horizon is proportional to the \emph{change} in that area. This is a natural generalization of the area law itself. We will argue that it is then hard to associate this quantity with a well-defined entropy of any particular local system.

\subsection{Formulation of thermodynamic gravity}\label{sec:TGformulation}

\begin{figure}[t]
\begin{center}
\includegraphics[width=.75\textwidth]{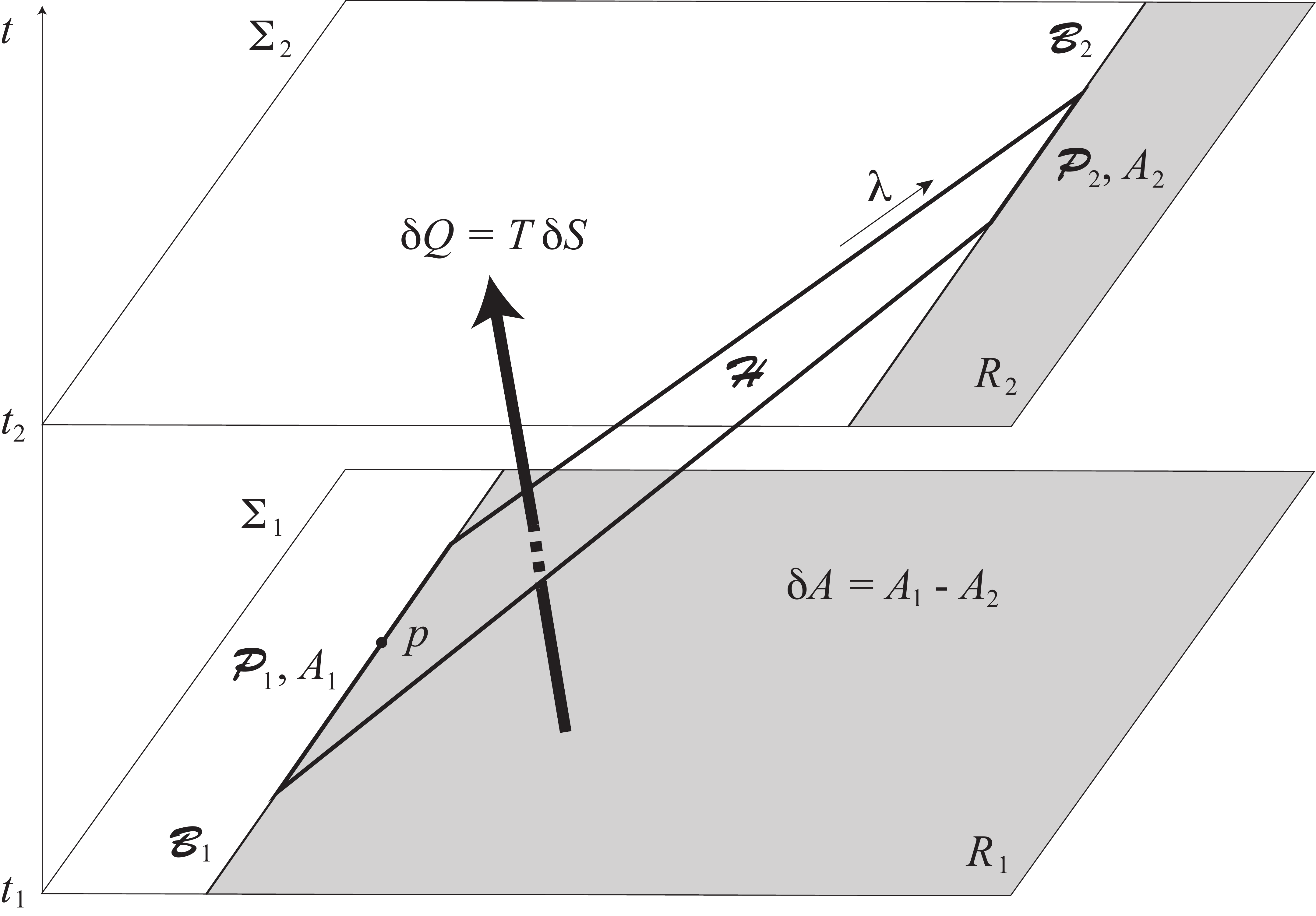}
\end{center}
\caption{Spacetime diagram of the flux through a segment $\mathcal{H}$ of a lightsheet. Starting with some point $p$, we fix an approximately-flat spacelike surface $\mathcal{P}_1 \ni p$ and a boundaryless surface $\mathcal{B}_1 \supset \mathcal{P}_1$. Consider future-directed inward null geodesics orthogonal to $\mathcal{P}_1$, with affine parameter $\lambda$. Flowing along the geodesics by some fixed parameter value, the area $A_1$ of the initial surface element $\mathcal{P}_1$ evolves into a new area $A_2$, which we can use to define the area decrement $\delta \mathcal{A} \equiv A_1-A_2$.
An amount of heat $\delta Q$ passes through $\mathcal H$, which by the Clausius relation is equal to $T\delta S$.
The regions $R_1$ and $R_2$ denote the parts of the spacelike hypersurfaces $\Sigma_1$ and $\Sigma_2$ that lie inside the spacelike codimension-two surfaces $\mathcal{B}_1$ and $\mathcal{B}_2$, respectively.}
\label{lightsheet-fig}
\end{figure}

Consider an arbitrary spacetime and identify some point $p$. 
Restrict to a sufficiently small region such that we can define a spacelike foliation with respect to a time coordinate $t$. 
Our point $p$ is located at time coordinate $t_1$ on a spacelike codimension-one hypersurface $\Sigma_1$. 
Choose a codimension-two approximately-flat spacelike surface $\mathcal{P}_1$ containing $p$.
Approximate flatness means that the null congruences normal to $\mathcal{P}_1$ have  vanishing expansion $\theta$ and shear $\sigma_{\mu\nu}$ at $p$ to first order in the distance from $p$. 
Fix a closed orientable smooth spacelike codimension-two surface $\mathcal{B}_1$ containing $\mathcal{P}_1$ and choose a future-directed inward null direction normal to $\mathcal{B}_1$, which defines a null congruence originating from $\mathcal{B}_1$.
Denote the spacelike region of $\Sigma_1$ that lies inside $\mathcal{B}_1$ by $R_1$.
Choose an affine parameter $\lambda$ along the congruence, with tangent vector $k^\mu = (\mathrm{d}/\mathrm{d}\lambda)^\mu$, letting $\lambda$ equal zero at $p$ and increase toward the future. 
Points in the congruence make up the ``lightsheet'' $\mathcal{H}$ emanating from $\mathcal{P}_1$. 
At a not-much-later time $t_2$, the intersection of the null congruence from $\mathcal{B}_1$ with a spacelike hypersurface $\Sigma_2$ defines a spacelike codimension-two surface $\mathcal{B}_2$, such that $\mathcal{P}_1$ evolves to $\mathcal{P}_2$.
The region inside $\mathcal{B}_2$ is denoted by $R_2$.
The setup is portrayed in \Fig{lightsheet-fig}.

The lightsheet $\mathcal{H}$ is a horizon in the sense that it serves as a local Rindler horizon for appropriately accelerating observers.
(In \Ref{Jacobson}, the construction was formulated over the past horizon instead of the future horizon, but this distinction makes no difference to our arguments.) 
We can define an approximate boost Killing vector $\chi^\mu=\kappa \lambda k^\mu$, where $\kappa$ is the acceleration of the associated Rindler trajectory. 
The surface element for the local Rindler horizon is $\mathrm{d} \Sigma^\mu = k^\mu \mathrm{d}\lambda\,\mathrm{d}\mathcal{A}$, where $\mathrm{d}\mathcal{A}$ is the codimension-two spacelike cross-sectional area element. 
This can be used to define a heat flux across the lightsheet
\be
\delta Q \equiv \int_{\mathcal H} T_{\mu\nu}\chi^\mu \mathrm{d}\Sigma^\nu = \kappa \int_{\mathcal H} T_{\mu\nu}  k^\mu k^\nu \lambda \, \mathrm{d}\lambda\,\mathrm{d}\mathcal{A}.\label{eq:dQ}
\ee
Viewing our system as the set of degrees of freedom on $R_1$ in \Fig{lightsheet-fig}, $\delta Q$ defines the heat leaving the system through $\mathcal{H}$. 
The temperature associated with this process is just the Unruh temperature \cite{Unruh} for the Rindler trajectory, $T=\hbar\kappa/2\pi$. 

The area decrement of the lightsheet as $\delta Q$ flows through it is
\be
\delta\mathcal{A} \equiv A_1 - A_2 = - \int_{\mathcal H} \theta\, \mathrm{d}\lambda\,\mathrm{d}\mathcal{A},\label{eq:dA}
\ee 
where $A_1$ is the initial area of the codimension-two surface $\mathcal{P}_1$, $A_1 = \int_{\mathcal{P}_1} \mathrm{d}\mathcal{A}$, and $A_2$ is the area of the codimension-two surface $\mathcal{P}_2$ at the other end of $\mathcal{H}$.
The expansion $\theta$ is defined to be $\theta \equiv \nabla_\mu k^\mu$.
Carefully treating the range of integration of $\lambda$ will play an important role in our discussion in \Sec{sec:vonNeumann}. 

Having made these preliminary definitions, we are ready to state the assumptions of thermodynamic gravity. They are the following:
\begin{enumerate}
\item {\bf Clausius relation.} There exists an entropy change $\delta S$ associated with the flow of heat through the lightsheet $\mathcal{H}$, which in local thermodynamic equilibrium is given by
\be
\delta S = \delta Q/T.\label{eq:Clausius}
\ee

\item {\bf Local holography.} For any lightsheet $\mathcal{H}$ of the form shown in \Fig{lightsheet-fig}, the entropy change $\delta S$ is proportional to the change in area $\delta\mathcal{A}$ with some universal constant $\eta$,
\be 
\delta S = \eta\,\delta\mathcal{A}.\label{eq:localholography}
\ee
\end{enumerate}

Note that the use of the Clausius relation \eqref{eq:Clausius} implies that the entropy $\delta S$ under consideration should correspond to some notion of entropy for a system that can be locally defined. The local holography assumption, meanwhile, is motivated by black hole thermodynamics, upon which entropic gravity is based. Therefore, we should expect that $\eta$ is the same coefficient as in the Bekenstein-Hawking formula \cite{HawkingBHThermo}, $1/4G\hbar$. Were we to find that $\eta\neq 1/4G\hbar$ is required for consistency with \Eqs{eq:Clausius}{eq:localholography} and Einstein's equation, this would undermine the original motivation for TG. This is the problem we will uncover in \Sec{sec:vonNeumann}.

Putting these assumptions together, we can derive Einstein's equation. First, from the Raychaudhuri equation,
\be
\frac{\mathrm{d}\theta}{\mathrm{d}\lambda} = -\frac{1}{D-2}\theta^2 - \sigma_{\mu\nu}\sigma^{\mu\nu} - R_{\mu\nu}k^\mu k^\nu,\label{eq:Raychaudhuri}
\ee
which is just a geometric statement in terms of the expansion, shear, and Ricci tensor $R_{\mu\nu}$,
\Ref{Jacobson} writes $\theta=-\lambda R_{\mu\nu}k^\mu k^\nu$ for a small segment of the lightsheet and inserts this result into \Eq{eq:dA} to obtain
\be
\delta\mathcal{A} =   \int_{\mathcal H}  R_{\mu\nu}k^\mu k^\nu\lambda\, \mathrm{d}\lambda\,\mathrm{d}\mathcal{A}.
\ee
Using local holography \eqref{eq:localholography} and the Clausius relation \eqref{eq:Clausius} to equate this to $\delta Q/T$, one can invoke the freedom in the choice of $k^\mu$ to equate the integrands, obtaining
\be
\eta (R_{\mu\nu}+f\,g_{\mu\nu}) =  \frac{2\pi}{\hbar} T_{\mu\nu} 
\ee
for some scalar quantity $f$. Since we must have $\nabla^\mu T_{\mu\nu} = 0$ for energy-momentum conservation, but $\nabla^\mu R_{\mu\nu}=\nabla_\nu R/2$ by the Bianchi identity, $f$ can be identified, yielding Einstein's equation,
\be
R_{\mu\nu} - \frac{1}{2}R\,g_{\mu\nu} + \Lambda g_{\mu\nu} = \frac{2\pi}{\hbar\eta}T_{\mu\nu} = 8\pi G \, T_{\mu\nu},\label{eq:Einstein}
\ee
where $\Lambda$ is the cosmological constant. We find that $\eta$ must indeed be equal to $1/4G\hbar$, as expected for consistency with the Bekenstein-Hawking formula.

We see that the assumptions of the Clausius relation \eqref{eq:Clausius} and local holography \eqref{eq:localholography} are, together, sufficient to derive Einstein's equation, at least up to a normalization. Less clear is the nature of the quantity $\delta S$ -- in particular, precisely what this is supposed to be the entropy \emph{of}. Formally, the only role of $\delta S$ in this derivation is to motivate equating $\eta\,\delta\mathcal{A}$ with $\delta Q/T$; once that happens, $\delta S$ disappears from the discussion. But if we were simply to assume $\eta\,\delta\mathcal{A} = \delta Q/T$ from the start, that would be tantamount to assuming Einstein's equation. The substantive content of TG, therefore, rests on the existence of a consistent and well-defined local construction for the entropy $\delta S$ associated with lightsheet segments anywhere in spacetime. We now turn to an investigation of what that construction might be.

\subsection{Entanglement entropy of a null region}\label{sec:vonNeumann} 

Some form of the von~Neumann entanglement entropy is a natural 
candidate for the quantity $\delta S$ that plays a crucial role in TG. We first need to specify the precise system whose entanglement entropy we are calculating.
Factors of Hilbert space are usually associated with regions of spacelike surfaces, but local holography refers to the entropy associated with part of a null surface. The simplest option would be to introduce some spacelike slicing, zoom in on a small neighborhood so that the spacetime looks approximately static, and compute the von~Neumann entropy on the small spacelike region; subsequently, one could enforce local holography on the small lightsheet through which the orthogonal timelike congruence originating from the small spacelike region passes. However, this prescription does not prove suitable: while the von~Neumann entropy is subextensive, energy-momentum is extensive. That is, considering two adjacent regions $A$ and $B$, we have $S_{AB} \leq S_A + S_B$, with strict inequality if $A$ and $B$ are entangled; however, the masses of $A$ and $B$, and hence the concomitant first-order area decrements of a lightsheet passing through them, add linearly. Thus, the use of the von~Neumann entropy on spacelike surfaces cannot provide a consistent formulation of thermodynamic gravity. 

\begin{figure}[t]
\begin{center}
\includegraphics[width=.65\textwidth]{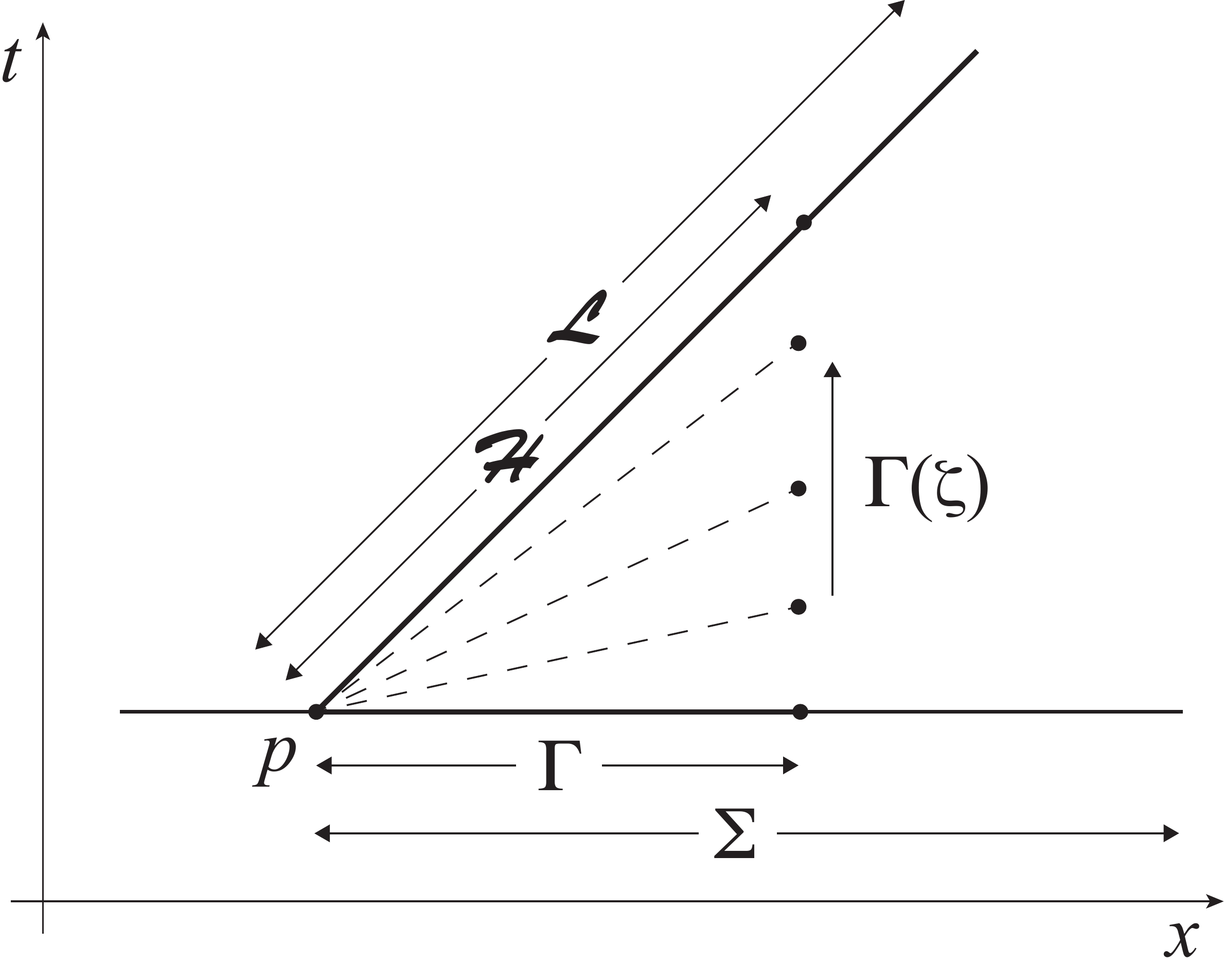}
\end{center}
\caption{A finite lightsheet $\mathcal{H}$ considered as the null limit of a parametrized collection of spacelike regions $\Gamma(\zeta)$. The large spacelike region $\Sigma$ maps to the large null surface $\mathcal{L}$. The affine parameter generating $\mathcal{H}$ runs from $0$ to $\epsilon$. }
\label{entropy-null-fig}
\end{figure}

We therefore turn to the null limit. Consider a spacelike region $\Sigma$, with a point $p \in \mathcal{P}_1$ on its boundary, as shown in \Fig{entropy-null-fig}.
It contains a smaller spacelike region $\Gamma$ with $p$ also on its boundary.
The large null surface to the future of $\Sigma$ is labeled $\mathcal{L}$ and a small lightsheet $\mathcal{H}$, as defined in \Sec{sec:TGformulation}, can be thought of as the null limit of a series of spacelike regions $\Gamma(\zeta)$.
There are then two different ways to associate an entropy with $\mathcal H$: {\it i}) the entanglement entropy associated with the region itself and {\it ii}) the difference in entanglement entropies between those of the large null surfaces $\mathcal{L}$ and $\mathcal{L}-\mathcal{H}$, which emanate from $\mathcal{P}_1$ and $\mathcal{P}_2$, respectively.
We will consider each possibility in turn.

Let us first see whether the entropy appearing in TG could be the entanglement entropy associated with the region $\mathcal{H}$. Let $\rho_{\Sigma}$ be the density matrix of the system on the spacelike region $\Sigma$ and let $\sigma_{\Sigma}$ be the vacuum density matrix.  Let $\sigma_\Gamma \equiv\Tr_{\Sigma-\Gamma}\sigma_\Sigma$ and $\rho_\Gamma  \equiv\Tr_{\Sigma-\Gamma}\rho_\Sigma$. We are immediately forced to identify some way to regulate the von~Neumann entropy, which naively diverges. Consider the vacuum von~Neumann entropy in the null limit, $\lim_{\Gamma\rightarrow\mathcal{H}} S(\sigma_\Gamma)$. If we simply impose an ultraviolet cutoff, the entanglement entropy $S(\sigma_\Gamma)$ associated with a vacuum region is still large \cite{Bombelli,Srednicki}, going as ${\cal A}/\epsilon^2$, where ${\cal A}$ is the area of the boundary of $\Gamma$ and $\epsilon$ is the cutoff length. By the local holography postulate \eqref{eq:localholography}, we must have $\delta S = \eta\, \delta{\cal A}$, where $\delta{\cal A}$ is the area decrement along $\mathcal{H}$, which must vanish in the Minkowski vacuum. While the details of a UV cutoff may have bearing on the renormalization of Newton's constant (see \Ref{Anber:2011ut} and references therein) and therefore of $\eta$, no such effect could reconcile a finite value of $\delta S$ with an exactly vanishing $\delta \mathcal{A}$. Thus, we cannot use the UV-regulated von~Neumann entropy in the null limit as $\delta S$ in entropic gravity, since doing so would require violation of either the postulate of local holography or flatness of the vacuum spacetime. We must therefore adopt the prescription of Casini \cite{Casini}, subtracting the entanglement entropy associated with the vacuum as in \Eq{eq:Casini}, producing the appropriate regulated version of the von~Neumann entropy that vanishes in vacuum.

 We compute the Casini entropy $\Delta S_\Gamma$ of the small spacelike region as the difference of the von~Neumann entropies for $\rho_\Gamma$ and $\sigma_\Gamma$ as in \Eq{eq:Casini}, $\Delta S_\Gamma \equiv  S(\rho_\Gamma) - S(\sigma_\Gamma)$, and then take the null limit to define the entropy on the small lightsheet, $\Delta S_\mathcal{H} \equiv \lim_{\Gamma\rightarrow {\mathcal{H}}}\; \Delta S_\Gamma$. Next, let us define a modular Hamiltonian $K_\Gamma$ on $\Gamma$ via
\be
\sigma_\Gamma \equiv \frac{e^{-K_\Gamma}}{\Tr\,e^{-K_\Gamma}}
\ee
and use this to define $\Delta K_\Gamma$ as in \Eq{eq:DeltaK}. Despite the nonlocality of $K$, the modular energy becomes more tractable in the null limit, $\Delta K_\mathcal{H} \equiv\lim_{\Gamma\rightarrow {\mathcal{H}}}\Delta K_\Gamma$, as we saw in \Sec{sec:HGjustify}.

It was shown in Refs.~\cite{Bousso1,Bousso2} for interacting quantum field theories that $\Delta S_\Gamma$ and $\Delta K_\Gamma$ become equal as the null limit is taken and, in particular,
\be 
\Delta S_\mathcal{H} = \Delta K_\mathcal{H} = \frac{2\pi}{\hbar}\int \mathrm{d}\mathcal{A}\int_0^\epsilon \mathrm{d}\lambda\,g(\lambda,\epsilon)T_{\mu\nu}k^\mu k^\nu,\label{eq:DeltaS}
\ee
where $g(\lambda,\epsilon)$ is a real function whose precise values depend on the interacting quantum field theory being considered. Note that $g(\lambda,\epsilon)$ is not automatically theory-independent, as in the causal diamond case: the causal diamond was related by a global conformal transformation \eqref{eq:conformaltransformation} to the Rindler wedge, while this is not so for the lightsheet $\mathcal{H}$. However, \Ref{Bousso2} showed that $g(\lambda,\epsilon)$ is computable in particular cases and moreover satisfies certain general properties for all interacting quantum field theories, which will be sufficient for our purposes.

The function $g(\lambda,\epsilon)$, whose properties we discuss in detail below, plays a crucial role here. Equations analogous to \Eq{eq:DeltaS} appear as expressions for the heat transfer in \Refs{Jacobson}{Non-equilibrium}, but with $g(\lambda,\epsilon)$ replaced simply by $\lambda$. [This similarity suggests that we should view \Eq{eq:DeltaS} as corresponding to the Clausius relation, indicating that this formulation of the entropy is appropriate for application to TG.] For the Rindler Hamiltonian, which inspires this form, $\lambda$ is perfectly appropriate for a semi-infinite lightsheet, but we are now computing the entropy for the \emph{finite} segment of lightsheet ${\cal H}$, for which the entropy takes the form of \Eq{eq:DeltaS},  as shown in Refs.~\cite{Bousso1,Bousso2}. That makes all the difference: $g(\lambda,\epsilon)$ initially increases as $\lambda$, but then decreases as $\epsilon-\lambda$ at the other end of the segment. As a result, the integral in \Eq{eq:DeltaS} differs from the Rindler Hamiltonian by a theory-dependent constant factor of order unity. This discrepancy implies that we cannot simultaneously choose our normalization so as to correctly recover Newton's constant in both Einstein's equation and in the area-entropy formula.

\begin{figure}[t]
\begin{center}
\includegraphics[width=.6\textwidth]{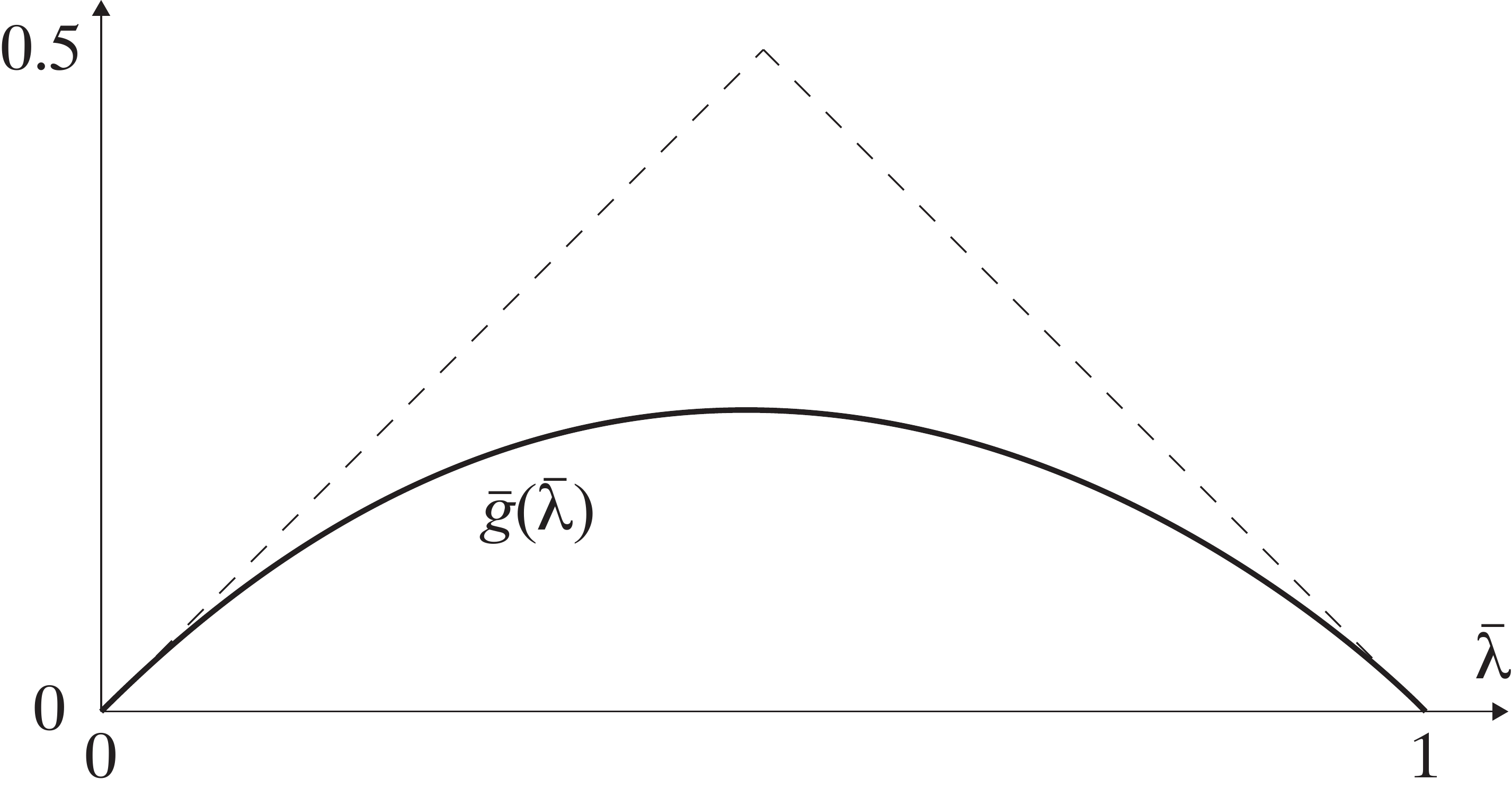}
\end{center}
\caption{Schematic form of the function $\bar g(\bar \lambda)$, proportional to the $g(\lambda,\epsilon)$ used in the expression for the null Casini entropy in \Eq{eq:DeltaS}. It is symmetric between $\bar\lambda=0$ and $\bar\lambda=1$, with slope between $1$ and $-1$ and a negative second derivative everywhere.}
\label{g-of-lambda-fig}
\end{figure}

Reference~\cite{Bousso2} derived a number of properties that the function $g(\lambda,\epsilon)$ appearing in \Eq{eq:DeltaS} must obey, amounting essentially to the requirement that it have the form illustrated in \Fig{g-of-lambda-fig}. More specifically, defining $\bar{\lambda}\equiv\lambda/\epsilon \in [0,1]$, we have $g(\lambda,\epsilon)=\epsilon \bar{g}(\bar{\lambda})$, with $\bar{g}(\bar{\lambda}) = \bar{g}(1-\bar{\lambda})$, and
\be
\begin{aligned}
\bar{g}(\bar{\lambda})&\rightarrow \bar{\lambda}&{\rm for}\;\bar{\lambda}\rightarrow 0,\\
\bar{g}(\bar{\lambda})&\rightarrow 1-\bar{\lambda}&{\rm for}\;\bar{\lambda}\rightarrow 1.\label{eq:gprops3}
\end{aligned}
\ee
Putting together the required properties of $\bar{g}$, \Ref{Bousso2} showed that $\left|{\rm d}\bar{g}/{\rm d}\bar{\lambda}\right| \leq 1$.
Note in particular that the integral $\int_0^1 \mathrm{d}\bar\lambda\, \bar{g}(\bar\lambda)$ is less than $1/4$.

Now let us consider the area variation of $\mathcal{H}$. As in \Ref{Jacobson}, we can choose $\mathcal{H}$ such that $\theta$ and $\sigma_{\mu\nu}$ vanish at first order near $p$. We can evaluate the change in the cross-sectional area of $\mathcal{H}$ by integrating the Raychaudhuri equation \eqref{eq:Raychaudhuri} for a finite lightsheet, keeping careful track of the ranges of integration. We find that the area decrement along $\mathcal{H}$ is
\be 
\Delta \mathcal{A} = -\int \mathrm{d}\mathcal{A}\int_0^\epsilon \mathrm{d}\lambda\, \theta(\lambda) 
=\int \mathrm{d}\mathcal{A}\int_0^\epsilon \mathrm{d}\lambda\, \int_0^\lambda \mathrm{d}\hat{\lambda}\, R_{\mu\nu}(\hat{\lambda})\hat{k}^\mu \hat{k}^\nu.\label{eq:DeltaA}
\ee

We can now test whether the null Casini entropy $\Delta S_\mathcal{H}$, which is the regularized von~Neumann entropy from \Eq{eq:Casini} evaluated in the null limit, can be the basis of a consistent formulation of TG. First, we need only consider the limit of a very small lightsheet, since we wish only to recover the local equations of motion, i.e., Einstein's equation. That is, we can take $\Delta \mathcal{A}$ and $\Delta S_\mathcal{H}$ in \Eqs{eq:DeltaA}{eq:DeltaS} in the limit of very small $\epsilon$ and cross-sectional area $\mathcal{A}$ to define $\delta \mathcal{A}$ and $\delta S$ for use in the assumption of local holography in \Eq{eq:localholography}. From \Eq{eq:DeltaA}, we have
\be
\delta \mathcal{A} \equiv \lim_{\epsilon\rightarrow {\rm small}} \lim_{\mathcal{A}\rightarrow{\rm small}} \Delta \mathcal{A} = \frac{1}{2}\epsilon^2 \mathcal{A}R_{\mu\nu}(p) k^\mu k^\nu,\label{eq:dAepsilon}
\ee
where we used the fact that in the limit of a small lightsheet the Ricci tensor could be taken to be a constant evaluated at $p$. Similarly, using \Eq{eq:DeltaS}, we find for the entropy that
\be
\delta S \equiv \lim_{\epsilon\rightarrow {\rm small}} \lim_{\mathcal{A}\rightarrow{\rm small}} \Delta S_\mathcal{H} = \frac{2\pi}{\hbar}\epsilon^2 \mathcal{A} T_{\mu\nu}(p) k^\mu k^\nu \int_0^1 \mathrm{d}\bar{\lambda}\,\bar{g}(\bar{\lambda}).
\label{eq:dSepsilon}
\ee

Local holography posits that $\delta S=\eta\,\delta \mathcal{A}$ for some constant $\eta$. For consistency with the Bekenstein-Hawking formula, we expect $\eta$ to equal $1/4G\hbar$, but for now we will keep it undetermined. 
Setting \Eq{eq:dAepsilon} proportional to \Eq{eq:dSepsilon} implies
\be
\left[\frac{4\pi}{\hbar\eta} \int_0^1 \mathrm{d}\bar{\lambda}\,\bar{g}(\bar{\lambda})\right] T_{\mu\nu}(p) k^\mu k^\nu= R_{\mu\nu}(p) k^\mu k^\nu.\label{eq:T=R}
\ee
Let us write $\eta = 1/4G_\mathrm{S}\hbar$ and write Newton's constant in Einstein's equation as $G_\mathrm{N}$. Then requiring consistency of \Eq{eq:T=R} with Einstein's equation and rearranging, we have
\be
G_\mathrm{S} = \frac{G_\mathrm{N}}{2\int_0^1 \mathrm{d}\bar{\lambda}\,\bar{g}(\bar{\lambda})} \geq 2 G_\mathrm{N},\label{eq:Gs}
\ee
noting, as we previously observed, that the integral over $\bar{g}(\bar{\lambda})$ is less than $1/4$. That is, in terms of the constant in Einstein's equation, we have
\be
\eta \leq \frac{1}{8G_\mathrm{N}\hbar}. \label{eq:etabound}
\ee
This is inconsistent, by an order-unity factor, with the area-entropy coefficient from black hole thermodynamics, which would be $\eta = 1/4G_\mathrm{N}\hbar$.
So we see that, while thermodynamic gravity is motivated by the area-entropy equivalence for black holes, enforcing $\delta S = \delta \mathcal{A}/4G\hbar$ would lead to the wrong constant in Einstein's equation. Moreover, this constant appears in a theory-dependent way via the function $\bar{g}$. On the other hand, one could insist that the correct coefficient be obtained in Einstein's equation. By \Eq{eq:Gs}, this would require $\delta S =\delta \mathcal{A} \int_0^1 \mathrm{d}\bar{\lambda}\, \bar{g}(\bar{\lambda}) / 2G\hbar$, which would constitute a theory-dependent modification of the local holography postulate with a coefficient that now no longer corresponds to the area-entropy relation from black hole thermodynamics. In other words, one could require that Einstein's equation and the $1/4G\hbar$ coefficient in the local holography postulate have Newton's constants that differ by the order-unity factor given in \Eq{eq:Gs}. A question for future work on TG would then be the identification of a justification, independent of Einstein's equation, of why the local holography postulate must take precisely this modified form.

The reason for the inconsistency of Einstein's equation and the expected area-entropy ratio in the formulation of TG we have considered here stems from the fact that, despite the similarity between \Eqs{eq:dQ}{eq:DeltaS}, there is a crucial factor-of-$g$ difference. In \Ref{Jacobson}, the heat transfer was taken to be given by the Rindler form~\eqref{eq:dQ}, where $g$ is just $\lambda$; interpreted as a modular Hamiltonian, this is the appropriate form for a semi-infinite lightsheet. However, only finite lightsheets \cite{Jacobson,Non-equilibrium} can be considered in the formulation of TG, so that $\theta$ and $\sigma_{\mu\nu}$ remain subdominant in the Raychaudhuri equation. 

There is an important distinction between the formulation of TG here and the causal-diamond derivation of HG in the previous section. The transformation \eqref{eq:conformaltransformation} that brings a Rindler wedge to the causal diamond is a true conformal transformation for the spacetime. In contrast, to bring a semi-infinite lightsheet to a finite segment requires a transformation $\lambda \rightarrow 1/\lambda$ that is conformal on two-dimensional subspaces, but not on the spacetime as a whole. For general theories (in particular, those that are not ultralocal), this leads to the need for the function $g(\lambda,\epsilon)$, which was not present for the causal-diamond formulation.

\subsection{Loopholes and alternatives}

A possible concern about this analysis might be that the Casini entropy \eqref{eq:DeltaS} is calculated in terms of the field-theoretic degrees of freedom alone. That is, one might imagine positing the existence of hidden, quantum-gravitational degrees of freedom that would provide additional entropy so that $\delta S$ equals $\delta \mathcal{A}/4G\hbar$, with the aim of getting both the correct coefficients in the area-entropy relation and in Einstein's equation. 

However, this proves to not be possible. The general form of the Casini entropy must be given by a relation of the form  \eqref{eq:DeltaS}, linear in the energy-momentum tensor, if we are to use $\delta S \propto \delta \mathcal{A}$ to derive Einstein's equation with $T_{\mu\nu}$ on the right-hand side. Positing new degrees of freedom can only affect the calculation of the theory-dependent coefficient $g(\lambda, \epsilon)$. But attaining $\eta=1/4G\hbar$ would require $|\mathrm{d}\bar{g}/\mathrm{d}\bar{\lambda}|$ to exceed unity. It is shown in \Ref{Bousso2} that this is impossible on very general grounds, regardless of any details about quantum field theory: exceeding this limit would violate strong subadditivity of von~Neumann entropy or monotonicity of quantum relative entropy.  Hence, positing non-field-theoretic degrees of freedom in the density matrix describing the lightsheet system is insufficient to simultaneously recover Einstein's equation and rectify the contradiction with the area-entropy formula we derived in \Eq{eq:etabound}. We are forced to conclude that the entropy in thermodynamic gravity cannot be the vacuum-subtracted von~Neumann (i.e., Casini) entropy of the lightsheet segment $\mathcal{H}$. 

An alternative tack for formulating TG would be to use the Casini entropy, but define the quantity $\delta S$ in a slightly different way.
Rather than associating it directly with the quantum state on the null region $\mathcal{H}$, we could let it be the \emph{difference} in Casini entropies between the large lightsheet $\mathcal{L}$ emanating from $p$ and the lightsheet with $\mathcal{H}$ removed, $\delta S = \Delta S_\mathcal{L} - \Delta S_\mathcal{L-H}$.
Note that this is in general a distinctly different quantity from that investigated above, since $\Delta S_\mathcal{H} \geq \Delta S_\mathcal{L} - \Delta S_\mathcal{L-H}$ by subadditivity.
For convenience, we will take $\mathcal{L}$ to be a semi-infinite null surface; because we are only interested in an entropy difference, the conclusions in this section are the same for any $\mathcal{L}$ much longer than $\mathcal{H}$. One might imagine that this alternate formulation, with semi-infinite lightsheets, would allow the Rindler form of the integrand in the expression for the entropy and possibly rescue thermodynamic gravity; however, this will prove to not be the case.

Let us specialize to spacetimes in which gravitational backreaction is small. (Including corrections to the Rindler Hamiltonian induced by spacetime curvature would only be relevant at higher order in Newton's constant.) Generalizing the arguments of \Ref{Bousso2} to semi-infinite null surfaces, with affine parameter $\hat{\lambda}$ going from $\lambda_0$ to infinity, we have
\be
\Delta S(\lambda_0) = \Delta K(\lambda_0) = \frac{2\pi}{\hbar}\int \mathrm{d}\mathcal{A}\int_{\lambda_0}^\infty (\hat\lambda-\lambda_0)T_{\mu\nu}\hat{k}^\mu \hat{k}^\nu \mathrm{d}\hat\lambda,\label{eq:DeltaSlambda0} 
\ee
where $\hat{k}^\mu=(\mathrm{d}/
\mathrm{d}\hat\lambda)^\mu$.
Then we can define the change in the null Casini entropy, $\delta S = \Delta S_\mathcal{L} - \Delta S_\mathcal{L-H} = \Delta S(\alpha) - \Delta S(\beta)$. Here, we have labeled the null regions by the value of the affine parameter from which they emanate, where in the $\hat\lambda$ parametrization, $\mathcal{H}$ is defined as $\hat\lambda \in [\alpha,\beta]$. 

However, this final formulation of the entropy as the null Casini entropy cannot be the correct definition of entropy in thermodynamic gravity. Let us define an affine parametrization that starts at $\lambda=1$ at $\hat\lambda=\alpha$, so $\lambda = \hat\lambda/\alpha$. Defining $k^\mu$ as the tangent four-vector to $\lambda$, $(\mathrm{d}/\mathrm{d}\lambda)^\mu$, we have
\be
\Delta S(\beta)=\frac{2\pi}{\hbar}\int_{\mathcal{A}(\beta\lambda^\prime/\alpha)} \mathrm{d}\mathcal{A}\int_1^\infty (\lambda^\prime - 1)T_{\mu\nu}(\beta\lambda^\prime/\alpha)k^{\prime\mu} k^{\prime\nu} \mathrm{d}\lambda^\prime,
\ee
where $\lambda^\prime=\alpha\lambda/\beta$ and $k^{\prime\mu}=(\mathrm{d}/\mathrm{d}\lambda^\prime)^\mu$.
We can make the approximation that $T_{\mu\nu}$ changes slowly with the affine parameter and that $\alpha$ and $\beta$ are close, so that $T_{\mu\nu}(\beta\lambda^\prime/\alpha)\simeq T_{\mu\nu}(\lambda^\prime)$. Further, we can take the cross-sectional area of the lightsheet to be small, so that $T_{\mu\nu}$ is approximately constant over the cross section at a fixed affine parameter. We thus have
\be
\delta S \simeq  \frac{2\pi}{\hbar}\int_1^\infty (\lambda - 1)[\mathcal{A}(\lambda)-\mathcal{A}(\beta\lambda/\alpha)]\,T_{\mu\nu}(\lambda)k^\mu k^\nu \mathrm{d}\lambda,\label{eq:deltaSCasini2}
\ee
where in the final line we dropped the primes, since $\lambda$ is a dummy variable. Now, from \Eq{eq:dAepsilon}, we have
\be
\mathcal{A}(\lambda)-\mathcal{A}(\beta\lambda/\alpha) \simeq \frac{1}{2}\left(\frac{\beta}{\alpha}-1\right)^2 \mathcal{A}(\lambda) R_{\mu\nu}(\lambda)k^\mu k^\nu,
\ee
Plugging this result into \Eq{eq:deltaSCasini2} and then substituting in Einstein's equation \eqref{eq:Einstein}, which implies $R_{\mu\nu} k^\mu k^\nu = 8\pi G T_{\mu\nu} k^\mu k^\nu$, we obtain
\be
\delta S = \frac{1}{8G \hbar}\left(\frac{\beta}{\alpha}-1\right)^2\int_1^\infty (\lambda-1)\mathcal{A}(\lambda) [R_{\mu\nu}(\lambda)k^\mu k^\nu]^2\label{eq:dSdiff} \mathrm{d}\lambda. 
\ee
We see that \Eq{eq:dSdiff} cannot be arranged in a form that looks like $\delta S = \eta\, \delta \mathcal{A}$ as in \Eq{eq:localholography}. In particular, \Eq{eq:dSdiff} is second-order rather than linear in the curvature and therefore in the area decrement $\delta \mathcal{A}$. Though \Eq{eq:DeltaSlambda0} looks similar to \Eq{eq:dQ}, one cannot naively conclude that the difference between the values of $\Delta S(\lambda_0)$ for $\lambda_0=\alpha$ versus $\beta$ in \Eq{eq:DeltaSlambda0} can be taken as simply an integral over $\lambda\in[\alpha,\beta]$; such an operation is not valid when the integrand itself has explicit dependence on its end points, as is the case in \Eq{eq:DeltaSlambda0}. We have found that by taking $\delta S$ to be the difference in the null Casini entropies of overlapping null surfaces, we obtain an expression \eqref{eq:dSdiff} for $\delta S$ that is fundamentally incompatible with the local holographic postulate \eqref{eq:localholography} that is one of the axioms of TG.

Hence, neither the null-limit Casini entropy of a small null region nor the difference in null-limit Casini entropies of two large null regions provides an acceptable definition of entropy in the thermodynamic formulation of entropic gravity.

\section{Conclusions}\label{sec:conclusions}

The idea that gravity can be thought of as an entropic force is an attractive one. In this paper we have distinguished between two different ways of implementing this idea: holographic gravity, which derives the Einstein equation from constraints on the boundary entanglement after varying over different states in the theory, and thermodynamic gravity, which relates the time evolution of a cross-sectional area to the entropy passing through a null surface in a specified spacetime. We argued that holographic gravity is a consistent formulation and indeed that recent work on the modular Hamiltonian in quantum field theory provides additional support for its underlying assumptions. The thermodynamic approach, on the other hand, seems to suffer from a difficulty in providing a self-consistent definition for what the appropriate entropy is going to be.

In the title of this work, we asked, ``What is the entropy in entropic gravity?'' We are now equipped to answer this question. In what we have called ``holographic gravity,'' the vacuum-subtracted von~Neumann entanglement entropy (the Casini entropy), evaluated on the null surfaces of the causal diamond, provides an appropriate formulation for an entropic treatment of gravitation. This can help guide further attempts to understand the underlying microscopic degrees of freedom giving rise to gravitation in general spacetime backgrounds.

\begin{center} 
 {\bf Acknowledgments}
 \end{center}

 \noindent 

We thank Cliff Cheung, Ted Jacobson, and Hirosi Ooguri for conversations that helped us considerably improve upon an earlier draft of this paper.
This research is funded in part by the Walter Burke Institute for Theoretical Physics at Caltech, by DOE grant DE-SC0011632, and by the Gordon and Betty Moore Foundation through Grant 776 to the Caltech Moore Center for Theoretical Cosmology and Physics. G.N.R. is supported by a Hertz Graduate Fellowship and a NSF Graduate Research Fellowship under Grant No. DGE-1144469.
\newpage

\bibliographystyle{utphys}
\bibliography{EntropicGravity}
\end{document}